\documentclass[a4paper]{article}
\usepackage{amsmath}
\usepackage{caption}
\begin{document}

\title{Unruh effect and entropy of Schwarzschild black hole}
\author{Teslyk M V \and Teslyk O M \and Zadorozhna L V}
\maketitle

\begin{abstract}
	We consider Unruh effect as an origin for Schwarzschild black hole entropy thus implying unitarian evolution of gravity. We simulate the black hole by set of Unruh horizons and estimate total entropy of the system. Dependence on mass and spin of the emitted particles is taken into account. The obtained results may be easily extended to any other intrinsic degrees of freedom of the particles. Unruh effect contribution to Schwarzschild black hole entropy is presented in exact analytical form.
\end{abstract}

\section{Introduction}
Bekenstein revealed in 1973 \cite{bhi_Bekenstein} that the black
hole (BH) entropy $S_{\rm BH}$ should be proportional to the event
horizon area $A$. A bit later black hole thermodynamics was
formulated \cite{bhi_4_laws} and in combination with
\cite{bhi_Hawking} determined that $S_{\rm BH}=A/4$ in Planck units.
Even for the solar mass scale $S_{{\rm BH}\odot}\sim10^{77}$, i.e.
BH has appeared to be a depository for a large amount of entropy.
Since then many approaches to explain such a challenge have been
proposed, see for example the recent reviews \cite{1804.10610,1807.05864} on the topic.

In 1996, Strominger and Vafa \cite{bhe_micro} proposed to consider
BH entropy with the help of string theory. It initiated a wide
investigation, as one can find in such reviews as
\cite{bhe_str,bhe_str_rev,bhe_str_rev_2}.

One more interesting approach is the loop quantum gravity where
entropy is being estimated via counting of BH microstates. It is
considered in such papers as
\cite{bhl_rad_spectr,bhe_Immirzi,bhl_su2,bhl_intertwiner}; for some
review one can read \cite{1201.6102}. In \cite{bhe_ev} authors
concluded that the BH radiation spectrum should become less entropic
as it evaporates. It leads to the possibility of information
read-out from the region behind the event horizon due to the
increasing contribution of quantum effects. Such a conclusion
implies information preservation with BH, see
\cite{bhi_Hawking_2,1403.7314}, and might interfere with \cite{a_bh_entr_en,a_bh_entr_en_small}.

BH event horizon separates the whole spacetime into accessible and
non-observable regions. Since any distant observer should trace out
all the degrees of freedom localized under the horizon, it is
supposed to be the generator for $S_{\rm BH}$
\cite{bhe_Srednicki,bhe_dof_rev}. At the same time the horizon can
be considered as a depository for the degrees of freedom where the
entropy originates from; it leads to holographic principle interpretation, see
\cite{bhef_holo,bhef_en,bhef_en_ext,1108.2650}.

In \cite{bhe_bw_unc} entropy divergence arising in 't Hooft's brick
wall model \cite{bhe_Hooft} was considered in the framework of the
uncertainty principle. The authors raise the question of similarity
between the entanglement and the statistical definitions of BH
entropy. In \cite{bhe_bw_cor}, higher order corrections within the
brick wall formalism for arbitrary spin have been found.

It is natural to assume that $S_{\rm BH}$ originates from
entanglement of the degrees of freedom being cut by the horizon. In
such a case quantum gravity should obey unitary evolution with no
information loss at all. The situation is
similar to partition of some quantum register into accessible and
non-accessible entangled parts. Having access to the part of the
code only (outside of the horizon), one meets with non-zeroth
entropy which may be reduced by reducing the inaccessible part
(inside the horizon). Such a restricted access can not be considered
as the one leading to entropy generation. For the other models
resulting in the unitary evolution of BH one can read
\cite{bhe_cond} and the review \cite{bhe_en_rev}.

Idea of entanglement entropy as a source for $S_{\rm
BH}$ has been used in \cite{a_bh_entr_en,a_bh_entr_en_small}.
However, the upper value for the entropy has appeared to be small: it did not exceed
1\% of $S_{\rm BH}$. Assuming that such a result was caused by
analysis of the spinless particles only, here we try to take spin degree of freedom into account. In this manuscript we
generalize entanglement entropy of outgoing radiation from the BH for both bosons and fermions.

This paper is organized as follows. In Section~\ref{sec:probability} we briefly discuss some important items from the probability theory and information entropy. Section~\ref{sec:Unruh} describes Unruh effect. Transition from Schwarzschild BH to a set of Unruh horizons is presented in Section~\ref{sec:model}. Estimation of the Unruh entropy and its asymptotics for Schwarzschild BH one may find in Section~\ref{sec:estimation}. Final conclusions and open questions one can read
in Section~\ref{sec:conclusion}.

\section{Probability and entropy}\label{sec:probability}
At first, consider some distribution $ \{X\} $ with unnormalized distribution probability $ d\left(x\right) $. In other words $ d\left(x\right) $ is a number of events when $ x $ is being observed. Shannon entropy $ H\left(X\right) $ for it may be written as
\begin{equation}\label{H(X)}
	H\left(X\right) = -\sum_{x}\frac{d\left(x\right)}{\#_{X}}\ln\frac{d\left(x\right)}{\#_{X}} = \ln\#_{X} - \frac{1}{\#_{X}}\sum_{x} d\left(x\right)\ln d\left(x\right),
\end{equation}
where $ \#_{X} = \sum_{x} d\left(x\right) $. $ H\left(X\right) $ defines amount of information we need to describe $ \{X\} $, i.e. amount of information we are lack of about the system, and therefore should deal with the distribution $ \{X\} $.

In case of joint distribution $ \{X,Y\} $ with unnormalized distribution probability $ d\left(x,y\right) $ the situation is similar to that in \eqref{H(X)}. Shannon entropy $ H\left(X,Y\right) $ for the distribution may be written also as
\begin{equation}\label{H(X,Y)}
	H\left(X,Y\right) = -\sum_{x,y}\frac{d\left(x,y\right)}{\#_{X,Y}}\ln\frac{d\left(x,y\right)}{\#_{X,Y}} = \ln\#_{X,Y} - \frac{1}{\#_{X,Y}}\sum_{x,y} d\left(x,y\right)\ln d\left(x,y\right),
\end{equation}
where $ \#_{X,Y} = \sum_{x,y} d\left(x,y\right) $.

At the same time, in the joint case one may define conditional probability $ d\left(x|y\right) $ as
\begin{equation}\label{d(x|y)}
	d\left(x|y\right) = \frac{d\left(x,y\right)}{d\left(y\right)}, \qquad d\left(y\right) = \sum_x d\left(x,y\right).
\end{equation}
It defines amount of events $ x $ from the set of events with $ y $. Using \eqref{H(X)}, Shannon entropy $ H\left(X|y\right) $ is then
\begin{equation}\label{H(X|y)}
	H\left(X|y\right) = \ln\#_{X|y} - \frac{1}{\#_{X|y}}\sum_{x}d\left(x|y\right).
\end{equation}

Finally, substituting \eqref{d(x|y)} and \eqref{H(X|y)} to \eqref{H(X,Y)} one obtains that
\begin{equation}\label{H conditional}
	H\left(X,Y\right) = H\left(Y\right) + \left\langle H\left(X|y\right)\right\rangle_{Y}
    				  = H\left(X\right) + \left\langle H\left(Y|x\right)\right\rangle_{X},
\end{equation}
where $ \left\langle\bullet\right\rangle_{Y}=\frac{1}{\#_Y}\sum_{y}d\left(y\right)\bullet $.

\section{Unruh entropy}\label{sec:Unruh}
Form here and below we use Planck units.

We consider Schwarzschild black hole of mass $ M $ and some quantum field surrounding it. The field is supposed to be
in a pure vacuum state $|0\rangle$ in the free-falling (Kruskal)
frame of reference (FR) and to have no influence on the background
metric (quasiclassical approach). Such a condition implies that the
field energy is negligible compared to BH's mass. Also we neglect total observer's energy and influence of the outgoing radiation on it for simplicity.

Any observer located at some fixed position right above the BH horizon should be described by non-inertial FR characterized by acceleration $ a = \left(4M\right)^{-1} $.

As it was demonstrated by Unruh in \cite{u_Unruh}, vacuum is
non-invariant with respect to FR. In the non-inertial FR appearance
of the horizon causes separation of the spacetime to the inside and
outside domains. As a result, the non-inertial observer detects some
radiation going out from the horizon, while the inertial one meets
the vacuum only. Then we have (see \cite{0903.0250,0908.3149} for
details)
\begin{equation}\label{|0> boson}
	|0\rangle = \sqrt{\frac{1-e^{-E/T}}{1-e^{-NE/T}}}\sum_{n=0}^{N-1} e^{-nE/2T}|n\rangle_{\rm in}|n\rangle_{\rm out},
\end{equation}
for bosons and
\begin{equation}\label{|0> fermion}
	|0\rangle = \frac{1}{\sqrt{1+e^{-E/T}}}\sum_{n=0}^{1} e^{-nE/2T}|n\rangle_{\rm in}|n\rangle_{\rm out},
\end{equation}
for fermions, where $ E $ is the energy of the quanta emitted at Unruh horizon with temperature $ T = \left(8\pi M\right)^{-1} = a\left(2\pi\right)^{-1} $. Parameter $ N $, as one may see from \eqref{|0> boson}, encodes maximum amount of quanta with energy $ E $ plus 1. The subscripts $_{\rm in},\,_{\rm out}$ denote the components of the field with respect to the horizon.

In the following we consider \eqref{|0> boson} only. Fermion case can be easily derived if one sets $ N=2 $, as it may be seen from comparison of \eqref{|0> boson} and \eqref{|0> fermion}.

Expression \eqref{|0> boson} is Schmidt decomposition. The outgoing radiation is described by the density matrix
\begin{equation}\label{rho}
	\rho_{\rm out} = {\rm Tr}_{\rm in}|0\rangle\langle0|
				   =\frac{1-e^{-E/T}}{1-e^{-NE/T}}\sum_{n=0}^{N-1}e^{-nE/T}|n\rangle_{\rm out}\langle n|
\end{equation}
where we have traced over the inaccessible degrees of freedom (in- modes). Thus, pure vacuum state from the Kruskal (inertial) FR has transformed into the mixed one in the Schwarzschild (non-inertial) FR. The only reason of the Unruh effect is the geometry: finiteness of the light speed leads to the appearance of the horizon dividing all the modes of the Hilbert space into the accessible (out-) and non-accessible (in-) ones. The complete state is obviously pure and follows the unitary evolution; but, since now one has limited access to it in the Kruskal FR, it looks like a non-unitarian evolution.

Density matrix \eqref{rho} describes conditional multiplicity distribution $ \{n|N,E/T\} $ at given $ N $ and $ E/T $. Its von Neumann entropy is defined as
\begin{eqnarray}\label{H_U(n|N,E/T)}
	H_{\rm U}\left(\rho_{\rm out}\right) = H\left(n|N, E/T\right)
										 = \sigma\left(E/T\right) - \sigma\left(NE/T\right),
\end{eqnarray}
where
\begin{equation}\label{sigma}
	\sigma\left(qE/T\right) = \frac{qE/T}{e^{qE/T}-1} - \ln\left(1-e^{-qE/T}\right).
\end{equation}

As one may notice, $ H\left(n|N, E/T\right) $ is an even function of $ E/T $. Its asymptotic behavior with respect to $ E/T $ is the following:
\begin{eqnarray}\label{H_U(n|N,E/T) limits}
	&&\lim_{E/T\to 0}H\left(n|N, E/T\right) = \ln N = \max\left(H\right),\nonumber\\
	&&\lim_{E/T\to\infty}H\left(n|N, E/T\right) = 0.
\end{eqnarray}

Expression \eqref{H_U(n|N,E/T)} is valid for some mode of the radiated field only, which is defined by energy $ E $ and parameter $ N $. However, since the emitted particles may have additional degrees of freedom (charge, spin etc.), one should take into account them too, which is the topic of the next Section.

\section{Model construction}\label{sec:model}
As we have mentioned above, \eqref{H_U(n|N,E/T)} does not take into account any other degrees of freedom but except multiplicity $ n $ and energy $ E $ as a parameter. As it was stated, fermions and bosons influence parameter $ N $ in different way: for fermions $ N \leq 2 $, while for bosons value of $ N $ is restricted by the quasiclassical approach limit and by energy conservation law only.

But, despite of the statistics, contribution of spin degrees of freedom has not taken into account. Since distribution $ \{n|N,E/T\} $ does not depend on spin, using \eqref{H conditional} one may write
\begin{equation}\label{H(s,n|N,E/T)}
	H\left(s,n|N,E/T\right) = \ln\left(2s+1\right) + H\left(n|N,E/T\right).
\end{equation}

One may argue that expression \eqref{H(s,n|N,E/T)} does not include contribution from any other degrees of freedom defining particles (such as charges, for example). But taking into account distribution over other quantum degrees of freedom $ \{Q\} $ with total dimension of the corresponding subspace $ N_{\rm Q} $ results in additional terms $ \ln N_{\rm Q} $ in the \emph{rhs} of \eqref{H(s,n|N,E/T)}. These terms are independent of energy $ E $ under the aforementioned quasiclassical approach, and hence one may simply substitute $ \ln\left(2s+1\right) \to \ln\left(2s+1\right) + \ln N_{\rm Q} $ in the following expressions. One may argue that spin distribution $ \{s\} $ should be included as any other $ \{Q\} $. However, we separated its contribution because of its influence on $ N $.

Entropy $ H_{\rm U}\left(\rho_{\rm out}\right) $ from \eqref{H_U(n|N,E/T)} describes outgoing radiation with axial symmetry. Unruh effect has only one fixed direction which is determined by unit vector $ \vec{a}/a $, and the outgoing radiation is being emitted by flat disk-shaped horizon of radius $ r = 2M $. Any of the Unruh horizons with different directions of $ \vec{a}/a $ generates the same amount of entropy $ H\left(s,n|N,E/T\right) $. Thus the total entropy equals to one source multiplied by amount of different distinguishable directions $ N_{\vec{a}/a} $, i.e.
\begin{equation}\label{H_BH(s,n|N,E/T)}
	H_{\rm BH}\left(s,n|N,E/T\right) = N_{\vec{a}/a}H\left(s,n|N,E/T\right).
\end{equation}

Number $ N_{\vec{a}/a} $ may be estimated via angular degrees of freedom in the following way.

Schwarzschild BH is an object with spherical symmetry. Any non-inertial observer measuring the outgoing radiation from the event horizon characterized by radius $ r $ with not of disk but of spherical shape. In order to take horizon's curvature into account, one should keep in mind that the emitted particle may be generated with non-zeroth value of angular momentum $ l $. It varies in range which may be defined as
\begin{equation}\label{l bounds}
	0 \leq \sqrt{l\left(l+1\right)} \leq \sqrt{L\left(L+1\right)} = rp = 2M\sqrt{E^2-m^2},
\end{equation}
where $ m $ is the rest mass of the particle.

Total amount of angular degrees of freedom $ N_{\vec{a}/a} $ is defined as the sum over all the values and projections of $ l $ available. In other words,
\begin{equation}\label{N_a}
	N_{\vec{a}/a} = \sum_{l=0}^{l=L}\sum_{-l}^{l} = \left(L+1\right)^2 = \frac{1}{4}\left(\sqrt{\frac{E^2-m^2}{4\pi^2T^2} + 1} + 1\right)^2,
\end{equation}
where \eqref{l bounds} has been used.

Estimation or measurement of energy distribution $ \{E\} $ for BH seems to be a difficult task to date. However, for large enough BH one may assume with high precision that this distribution is homogeneous, i.e. that $ H\left(E\right) \approx \ln N_E $. In other words, we assume probability of particle emission to be independent of energy $ E $.

Amount of energy states $ N_E $ depends on the energy precision $ \Delta_E $. From the uncertainty principle $ \Delta_E\Delta_t\approx 1 $ we obtain that
\begin{equation}\label{N_E}
	N_E  = \frac{1}{\Delta_E}\int_{m}^{1}{\rm d}E \approx \Delta_t\int_{m}^{1}{\rm d}E = \Delta_t\left(1-m\right),
\end{equation}
where $ \Delta_t $ is time interval during which we observe the BH radiation. It determines maximum number of particles to be observed also, i.e. $ N-1 $ per energy $ E $; the only restriction for it is that total energy emitted should obey the quasiclassical approach, see Section~\ref{sec:Unruh}. Also we assume that energy $ E $ of the particle is defined in range $ m \leq E \leq 1 $.

In the following we consider homogeneous emission probability with respect to energy. This is valid in case of quasiclassical approach only, when the BH mass is large compared to the energy emitted outside. Otherwise one is expected to take into account back-reaction as quantum gravity effects. From \eqref{H conditional} we obtain
\begin{eqnarray}\label{H_BH(s,n,N,E|T)}
	H_{\rm BH}\left(s,n,E|T\right) &\approx& \ln N_E + \frac{1}{N_E}\sum_{\{E\}}H_{\rm BH}\left(s,n|N,E/T\right)\nonumber\\
								   &\approx& \ln N_E + \frac{1}{1-m}\int_{m}^{1}N_{\vec{a}/a}H\left(s,n|N,E/T\right){\rm d}E,
\end{eqnarray}
where in the 2nd line we used \eqref{H_BH(s,n|N,E/T)} and replaced summation with integral over the energies. Parameter $ N $ is omitted because it is completely defined by the spin $ s $ and by the observation time $ \Delta_t $, see \eqref{N_E} and the text below.

%Dependence of the integrand on energy is completely determined by \eqref{N_a}, \eqref{H(s,n|N,E/T)} and \eqref{H_U(n|N,E/T)}.

Entropy $ H_{\rm BH}\left(s,n,E|T\right) $ defines particle emission, which is being described by joint distribution $ \{s,n,E|T\} $, from the BH horizon with temperature $ T $ due to Unruh effect.

\section{Entropy estimation}\label{sec:estimation}
In order to proceed we rewrite \eqref{H_BH(s,n,N,E|T)} using \eqref{N_a} and \eqref{H(s,n|N,E/T)} in the following form:
\begin{equation}\label{H_BH/S_BH}
	\frac{H_{\rm BH}\left(s,n,E|T\right)}{S_{\rm BH}} \approx 16\pi T^2\ln N_E + \varUpsilon_{\rm dof} + \varUpsilon_{\rm U},
\end{equation}
where $ S_{\rm BH} = \left(16\pi T^2\right)^{-1} $ is the BH entropy and the 1st term is defined by \eqref{N_E}.

The 2nd term $ \varUpsilon_{\rm dof} $ may be calculated analytically:
\begin{eqnarray}\label{varUpsilon dof}
	\varUpsilon_{\rm dof} &=& 4\pi T^2\frac{\ln\left(2s+1\right)}{1-m}\int_{m}^{1}\left(\sqrt{\frac{E^2-m^2}{4\pi^2T^2} + 1} + 1\right)^2{\rm d}E\nonumber\\
						  &=&  \Biggl(	  \frac{1+m-2m^2}{3\pi}
						  				+ 2T\frac{\sqrt{1+4\pi^2T^2-m^2}}{1-m}
						  				+ 4\pi T^2\frac{1-2m}{1-m}\nonumber\\
						  &&\quad		+ 2T\frac{4\pi^2T^2-m^2}{1-m}\ln\frac{1+\sqrt{1+4\pi^2T^2-m^2}}{2\pi T +m}
						  		\Biggr)\ln\left(2s+1\right).
\end{eqnarray}
It describes contribution of the intrinsic degrees of freedom (spin and charges, see the text below \eqref{H(s,n|N,E/T)}) of outgoing radiation to the total entropy. 

The 3rd term $ \varUpsilon_{\rm U} $, see \eqref{H_U(n|N,E/T)},
\begin{equation}\label{varUpsilon U 1}
	\varUpsilon_{\rm U} = \frac{4\pi T^2}{1-m}\int_{m}^{1}\left(\sqrt{\frac{E^2-m^2}{4\pi^2T^2} + 1}
						 + 1\right)^2\left[\sigma\left(E/T\right)-\sigma\left(NE/T\right)\right]{\rm d}E
\end{equation}
defines gravitational Unruh contribution to the total entropy and, as one may notice, contains incomplete Bose-Einstein integrals. It may be calculated in the following way. 

Rewriting \eqref{sigma} as
\begin{equation*}
	\sigma(qE/T) = \sum_{k=1}^{\infty}\left(qE/T + 1/k\right)e^{-kqE/T}
\end{equation*}
and using lower incomplete gamma functions $ \gamma\left(\nu,x\right) $
\begin{equation*}
\gamma\left(\nu,x\right) = \int_{0}^{x}t^{\nu-1}e^{-t}{\rm d}t = \left(\nu-1\right)!\left(1 - e^{-x}\sum_{j=0}^{\nu-1}\frac{x^j}{j!}\right),
\end{equation*}
%is the lower incomplete gamma function.
one may calculate the following integral as
\begin{eqnarray}\label{integral E^nu sigma}
\int_{m}^{1}\sigma\left(qE/T\right)E^\nu{\rm d}E = \frac{T^{\nu+1}}{q^{\nu+1}}\sum_{k=1}^{\infty}
													\left.\frac{\gamma\left(\nu+1,x\right)+\gamma\left(\nu+2,x\right)}{k^{\nu+2}}\right|_{x=kqm/T}^{x=kq/T}.
\end{eqnarray}
%\begin{eqnarray}\label{integral E^nu sigma}
%	\int_{m}^{1}\sigma\left(qE/T\right)E^\nu{\rm d}E 
%		&=& \frac{T^{\nu+1}}{q^{\nu+1}}\sum_{k=1}^{\infty}
%			 \left.\frac{\gamma\left(\nu+1,x\right)+\gamma\left(\nu+2,x\right)}{k^{\nu+2}}\right|_{x=kqm/T}^{x=kq/T}\\
%		&=& \frac{T^{\nu+1}}{q^{\nu+1}}\sum_{k=1}^{\infty}\left.\frac{e^{-x}}{k^{\nu+2}}\left[\nu!\left(\nu+2\right)\sum_{j=0}^{\nu}\frac{x^j}{j!}
%			 +x^{\nu+1}\right]\right|_{x=kq/T}^{x=kqm/T}.\nonumber
%\end{eqnarray}
Substituting \eqref{integral E^nu sigma} to \eqref{varUpsilon U 1} and using decomposition
\begin{equation*}
	\left(1+x\right)^{\alpha} = \sum_{q=0}^{\infty}\binom{\alpha}{q}x^q,\quad |x|<1,
\end{equation*}
one obtains that
\begin{eqnarray}\label{varupsilon U 2}
	\varUpsilon_{\rm U} &=& \Biggl[
									\left(8\pi^2 T^2-m^2\right)\sum_{k=1}^{\infty}\frac{\gamma\left(1,x\right) + \gamma\left(2,x\right)}{k^{2}}
										\left(\Bigl.\Bigr|_{x=km/T}^{x=k/T}	- \frac{1}{N}\Bigl.\Bigr|_{x=kNm/T}^{x=kN/T}\right)\nonumber\\
						&&\quad		+ T^2\sum_{k=1}^{\infty}\frac{\gamma\left(3,x\right) + \gamma\left(4,x\right)}{k^{4}}
										\left(\Bigl.\Bigr|_{x=km/T}^{x=k/T}	- \frac{1}{N^3}\Bigl.\Bigr|_{x=kNm/T}^{x=kN/T}\right)\nonumber\\
						&&\quad		+ 8\pi^2\sum_{n=0}^{\infty}\sum_{q=0}^{\infty}m^{2q}T^{2-2q}\left(-1\right)^{q}\binom{1/2}{n}\nonumber\\
						&&\qquad		\times\sum_{k=1}^{\infty}\begin{cases}
																 	A_{nqk}\left(N,m,T\right),	& 2\pi T > \sqrt{E^2-m^2}\\
																	B_{nqk}\left(N,m,T\right),	& 2\pi T < \sqrt{E^2-m^2}
																 \end{cases}\nonumber\\
						&&	\Biggr]\frac{T}{\pi\left(1-m\right)},
\end{eqnarray}
where
\begin{eqnarray*}
	A_{nqk}\left(N,m,T\right) &=& \binom{n}{q}\frac{\gamma\left(1+2n-2q,x\right) + \gamma\left(2+2n-2q,x\right)}{\left(2\pi\right)^{2n}k^{2+2n-2q}}\\
							  &&\times\left(\Bigl.\Bigr|_{x=km/T}^{x=k/T} - \frac{1}{N^{1+2n-2q}}\Bigl.\Bigr|_{x=kNm/T}^{x=kN/T}\right)
\end{eqnarray*}
and
\begin{eqnarray*}
	B_{nqk}\left(N,m,T\right) &=& \binom{1/2-n}{q}\left(2\pi\right)^{2n-1}
									\frac{\gamma\left(2-2n-2q,x\right) + \gamma\left(3-2n-2q,x\right)}{k^{3-2n-2q}}\\
							  &&\times\left(\Bigl.\Bigr|_{x=km/T}^{x=k/T} - \frac{1}{N^{2-2n-2q}}\Bigl.\Bigr|_{x=kNm/T}^{x=kN/T}\right).
\end{eqnarray*}

To sum up, we have calculated in analytical form Unruh entropy $ H_{\rm BH}\left(s,n,E|T\right) $ for the particles emitted at the Schwarzschild BH horizon, see \eqref{H_BH/S_BH}. The estimation takes into account spin and other (charges) degrees of freedom of the emitted particles. The entropy may be separated to 3 different terms that are responsible for different aspects of the Unruh radiation, see \eqref{N_E}, \eqref{varUpsilon dof} and \eqref{varupsilon U 2}.

In order to explore the extreme cases (large and small BHs) and the applicability of \eqref{H_BH/S_BH} we investigate asymptotics of $ H_{\rm BH}\left(s,n,E|T\right) $ below. 

In case of large BHs ($ T \to 0 $) both the 1st and the 3rd terms of \eqref{H_BH/S_BH} vanish, see \eqref{N_E} and \eqref{varupsilon U 2}. Thus one obtains that, see \eqref{varUpsilon dof},
\begin{eqnarray}\label{H_BH T -> 0}
	\lim\limits_{T \to 0}\frac{H_{\rm BH}\left(s,n,E|T\right)}{S_{\rm BH}}
		&=&		\lim\limits_{T \to 0}\varUpsilon_{\rm dof} = \frac{1+m-2m^2}{3\pi}\ln\left(2s+1\right)\nonumber\\
		&\geq&	\frac{\ln\left(2s+1\right)}{3\pi},
\end{eqnarray}
As one may notice, particles with non-zeroth $ m $ increase the entropy. However, the contribution is insignificant due to the fact that $ m \ll 1 $ for any elementary particles being known to date. Anyway, the lower bound for the contribution of Unruh entropy to $ S_{\rm BH} $ is non-negative and makes a rather significant contribution to $ S_{\rm BH} $, see Table~1. The term surviving the limit from \eqref{H_BH T -> 0} is the only one obeying the area law in the model proposed.
\begin{table}
	\begin{tabular}{|l|*{6}{c}|}
		\hline
		spin $ s $											& 0	& 1/2	& 1		& 3/2	& 2		& 5/2	\\ \hline
		$ \left(3\pi\right)^{-1}\ln\left(2s+1\right) $, \%	& 0	& 7.35	& 11.66	& 14.71 & 17.08	& 19.01	\\
		\hline
	\end{tabular}
	\caption{lower bounds for $ \lim\limits_{T \to 0}\frac{H_{\rm BH}\left(s,n,E|T\right)}{S_{\rm BH}} $ at different values of spin.}
\end{table}

The case $ T \neq 0 $ includes small BHs. As it follows from \eqref{H_BH/S_BH} (see \eqref{N_E}, \eqref{varUpsilon dof}, \eqref{varupsilon U 2}), Unruh entropy quickly rises up with $ T $ increasing. It reaches maximum at $ T = 1 $ that exceeds 1. Thus the proposed model is not valid in case of $ T \approx 1 $, as it has been stated at the very beginning of Section~\ref{sec:Unruh}, when we have restricted ourselves to the quasiclassical approach. For $ T \approx 1 $ one is expected to describe the entropy in terms of quantum gravity. 

As one may notice, entropy $ H_{\rm BH}\left(s,n,E|T\right) $ contains negative terms proportional to powers of $ m $, see \eqref{varUpsilon dof} and \eqref{varupsilon U 2}. Some of them survive in the case of large BHs even, see \eqref{H_BH T -> 0}. The terms may witness the presence of correlations in the outgoing radiation encoded by massive particles only. It may be interpreted as some information outflow from the BH, see for example \cite{bhi_Hawking_2}. For some more information on the topic one may see \cite{a_bh_entr_en}.

\section{Conclusions}\label{sec:conclusion}
In the paper we considered BH entropy $ S_{\rm BH} $ as the one resulting from the Unruh effect, when any non-inertial observer, while being located at the fixed distance from the BH horizon, is expected to deal with Unruh radiation. In order to do that we replaced the Schwarzschild BH by some set of Unruh horizons since they obey different symmetries (spherical and cylindrical respectfully). The approach is valid in case of quasiclassical approach only, when any back-reaction and quantum gravity effects are insignificant. The method allows to take into account all the intrinsic degrees of freedom of the emitted particles (spin $ s $, charge etc.), their mass $ m $ and multiplicity (defined by the parameter $ N $).

Considering Unruh effect as the origin for $ S_{\rm BH} $ one reduces the problem of BH entropy to be a geometrical one. From the information theory point of view, the BH entropy then is nothing more but the result of inaccessibility to the whole state, similar to entropy one is dealing with while reading only some part of the register. Therefore in our approach we imply that gravity obeys unitarian evolution, i.e. preserves the information.

Following the assumptions above we calculated entropy $ H_{\rm BH}\left(s,n,E|T\right) $, see \eqref{H_BH/S_BH}, in exact analytical form. As one may notice, it contains terms obeying the area law, see \eqref{varUpsilon dof}, that are the leading ones with respect to BH temperature $ T $. Surprisingly enough, these terms are not defined by the Unruh effect but by the intrinsic degrees of freedom only, see \eqref{varUpsilon dof} and \eqref{varupsilon U 2}.

Different spins give different contributions to $ H_{\rm BH}\left(s,n,E|T\right) $, see \eqref{varUpsilon dof}. Contribution to entropy for the leading term is significant and increases with spin, see Table~1.

Interestingly enough, $ H_{\rm BH}\left(s,n,E|T\right) $ contains negative terms proportional to powers of $ m $. They reduce the entropy of the outgoing radiation that may be interpreted as some information outflow from the BH. The effect does not vanish even in case $ T \to 0 $ (large BHs).

The approach has much in common with the one presented in \cite{bhe_en_short,bhe_en_2}. Compared to the papers, we take into account mass $ m $ of the emitted particles, their multiplicity and intrinsic degrees of freedom and present the result in exact analytical form.

In its current form the method can not explain all the entropy for BH, especially for small BHs ($ T \approx 1 $). The main problems are the assumption of homogeneous particle emission with respect to energy and neglect of back-reaction that both require taking into account quantum gravity effects. Thus we conclude that the problem is open still and deserves further investigation.

\bibliography{D:/Bibliography/Catalogue/bib}
\end{document}